# Study of a new link layer security scheme in a wireless sensor network

Nasrin Sultana[*]    Tanvir Ahmed[**]    Professor Dr. ABM Siddique Hossain [Ψ]

**Abstract — Security of wireless sensor network (WSN) is always considered a critical issue and has a number of considerations that separate them from traditional wireless sensor network. First, sensor devices are typically vulnerable to physical compromise. Second, they have significant power and processing constraints. Third, the most critical security issue is protecting the aggregate output of the system, even if individual nodes may be compromised. While a variety of security techniques are being developed and lots of researches are going on security fields. In this paper we have proposed a new technique to provide data authentication and privacy in faster, scalable and cost effective way.**

## 1. Introduction

Wireless sensor networks are application dependent and primarily designed for real-time collection and analysis of low level data in hostile environments. Popular wireless sensor network applications include wildlife monitoring, bushfire response, military command, intelligent communications, industrial quality control, observation of critical infrastructures, smart buildings, distributed robotics, traffic monitoring, examining human heart rates etc. Majority of the sensor network are deployed in hostile environments with active intelligent opposition. Hence security is a crucial issue for such type of applications.  One obvious example is battlefield applications where there is a pressing need for secrecy of location and resistance to subversion and destruction of the network [4]. The wireless medium is inherently less secure because its broadcast nature makes eavesdropping simpler. Any transmission can easily be intercepted, altered, or replayed by an adversary. The wireless medium allows an attacker to easily intercept valid packets and easily inject malicious ones.

[*]Lecturer, Department of Computer Science, American International University Bangladesh (AIUB) Kemal Ataturk Avenue, Dhaka, Bangladesh, Email: - nasrin@aiub.edu

[**]Lecturer, Department of Computer Science, American International University Bangladesh (AIUB) Kemal Ataturk Avenue, Dhaka, Bangladesh, Email:- tanvir@aiub.edu

[Ψ] Professor and Dean, Faculty of Engineering, American International University Bangladesh (AIUB) Kemal Ataturk Avenue, Dhaka, Bangladesh,,Email : siddique@aiub.edu

Encryption and authentication using cryptographic techniques makes a system significantly more secure against eavesdropping and other attacks. Encryption can be used to keep data secure from the adversary, and authentication can be used to safeguard against spurious data. In essence, these techniques attempt to ensure system-level confidentiality by protecting all links.

The security issues in WSN are more challenging than those in traditional wired networks. Most sensor networks actively monitor their surroundings, and it is often easy to deduce information other than the data monitored. Such unwanted information leakage often results in privacy breaches of the people in the environment. Moreover, the wireless communication employed by sensor networks facilitates eavesdropping and packet injection by an adversary. The combination of these factors demands security for sensor networks at the time of design to ensure operation safety, secrecy of sensitive data, and privacy for people in sensor environment [1]. Significant efforts and research have been undertaken to enhance security levels of wireless networks.

Security in sensor networks is complicated by the constrained capabilities of sensor node hardware and the properties of the deployment [1], [2] and [3].

Depending on application we are being motivated to propose efficient and effective block cipher architecture of link layer security in a wireless sensor network. Effective eavesdropping can be prevented in wireless sensor network with faster and cost effective way with a minimum good security like message/entity authentication or a combination of confidentiality and authentication. This new class of networks closely resembles the behavior of wireless networks. Nevertheless, they have a few unique differences; the principal one is the small size of Nodes constituting a WSN. Although smaller nodes make WSNs suitable for several existing and emerging applications related to information sensing, this also implies that the nodes have limited resources, like CPU speed, memory, battery, and radio interface. Because the nodes are resource constrained, they require network designs that can be customized for different types of application environments, thus placing significant demands on algorithm design, protocol specification, and technologies.

Sensor networks simplify the simultaneous collection and organization of data from multiple locations,

which may be unreachable, inhospitable, or even hostile environments. Merging wireless communications with sensor network capabilities enables rapid deployment and reduces the cost of the infrastructure. However, adopting wireless communications introduces a new set of challenges.

Sensor nodes are susceptible to physical capture, but because of their targeted low cost, tamper-resistant hardware are unlikely to prevail. Sensor nodes use wireless communication, which is particularly easy to eavesdrop on.

Attacker can easily inject malicious messages into the wireless network.

Anti-jamming techniques such as frequency- hopping spread spectrum and physical tamper proofing of nodes are generally impossible in a sensor network due to the requirements of greater design complexity and higher energy consumption.

The use of radio transmission, along with the constraints of small size, low cost, and limited energy, make WSNs more susceptible to denial-of-service attacks.

Ad-hoc networking topology of WSN facilitates attackers for different types of link attacks ranging from passive eavesdropping to active interfering.

Most current standard security protocols were designed for two-party settings and do not scale to a large number of participants.

There is a conflicting interest between minimization of resource consumption and maximization of security level.

Since sensor nodes usually have severely constrained applications. Thus, a promising approach is to use more efficient symmetric cryptographic alternatives. Instead, most security schemes make use of symmetric key cryptography. One thing required in either case is the use of keys for secure communication. Managing key distribution is not unique to WSNs, but again constraints such as small memory capacity make centralized keying techniques impossible.

In section 2 we discussed the related works. Our proposed model for a new link layer security scheme is described in section 3. We have a discussion in section 4. Finally the paper concludes with the idea for future advancement in section 5.

## 2. Related Works

TinySec [5] is link layer security architecture for wireless sensor networks, implemented for the TinyOS operating system. In order to overcome the processor, memory and energy constraints of sensor nodes TinySec leverages the inherent sensor network limitations, such as low bandwidth and relatively short lifetime for which the messages need to remain secure, to choose the parameters of the cryptographic primitives used. TinySec has two modes of operation: authenticated encryption (TinySec-AE) and authentication only (TinySec-Auth). With authenticated encryption, TinySec encrypts the data payload and authenticates the packet with a message authentication code (MAC). The MAC is computed over the encrypted data and the packet header. In authentication only mode, TinySec authenticates the entire packet with a MAC, but the data payload is not encrypted.

An important feature of TinySec is its ease of use and transparency, as many application developers will either implement the security features incorrectly or leave out any security entirely if the security API is difficult to use. TinySec solves this problem by integrating into TinyOS at a low level.

MiniSec [6] is a secure network layer protocol that claims to have lower energy consumption than TinySec while achieving a level of security which matches that of Zigbee. A major feature of MiniSec is that it uses offset codebook (OCB) mode as its block cipher mode of operation, which offers authenticated encryption with only one pass over the message data. Normally two passes are required for both secrecy and authentication. Another major benefit of using OCB mode is that the ciphertext is the same length as the plaintext, disregarding the additional fixed length tag, four bytes in MiniSec's case, so padding or ciphertext stealing is not necessary. Another primary feature MiniSec has over the other security suites mentioned here is strong replay protection without the transmission overhead of sending a large counter with each packet or the problems associated with synchronized counters if packets are dropped. To achieve this MiniSec has two modes of operation, one for unicast packets MiniSec-U and one for broadcast packets, MiniSec-B as explained.

Exploring security issues in wireless sensor networks, and in particular, we propose an efficient link layer security scheme inspired by TinySec [5]. To meet the desideratum of minimizing computation and communication overhead, our focus is the CBC-X mode Encryption/Decryption algorithm, which enables encryption/decryption and authentication of packets a One-pass operation. In particular, the main contributions of the paper are as follows:

We present an efficient link layer security scheme that attains confidentiality and authentication of packets in wireless sensor networks. Security services are provided transparently to the upper (link) layers of the protocol stack. We devise a CBC-X mode symmetric key mechanism to implement our link layer security scheme. Encryption/Decryption and authentication operations are combined as a one-pass Operation, which reduces half of the computational overhead of computing them separately.

The padding technique makes the scheme have no cipher text expanding for the transmitted data payload.

Hence it significantly reduces communication overhead.

Clearly, the CBC-X mode symmetric key mechanism in combination with the padding technique enables our scheme to substantially save power consumption of sensor nodes.

### 2.1 Problem statements

A wireless sensor network (WSN) is a network comprised a large number of sensors that are physically small Communicate wireless among each other and are deployed without prior knowledge of the network topology. Due to the limitation of their physical size, the sensors tend to have storage space, energy supply and communication bandwidth so limited that every possible means of reducing the usage of these resources is aggressively sought. Typical wireless sensor networks comprise of the wireless sensor nodes logically interconnected to each other, to realize some vital functionality. Wireless sensor nodes are characterized by severe constraints in power, computational resources, memory, and bandwidth and have small physical size with low power consumption. Since the Processing of the data is done on-the-fly, while being transmitted to the base station, the overall communication costs are reduced. Due to the multi-hop communication and the process demanding applications, the conventional end-to-end security mechanisms are not feasible for the WSN. Hence, the use of the standard end-to-end security protocols like SSH, SSL or IPSec in WSN environment is rejected. Instead, appropriate link layer security architecture, with low associated overhead is required. There are a number of research attempts that aim to do so.

### 3. Proposed Model

Cryptographic algorithms are an essential part of the security architecture of WSNs, using the most efficient and sufficiently secure algorithm is thus an effective means of conserving resources. By 'efficient' in this paper we mean requiring little storage and consuming less energy. The essential cryptographic primitives for WSNs are block ciphers, message authentication codes (MACs) and hash functions. Among these primitives, we are only concerned with block ciphers in this paper, because MACs can be constructed from block ciphers [7] where as hash functions are relatively cheap [8]. Meanwhile, public-key algorithms are well-known to be prohibitively expensive [9].

The most important parameters of a block cipher are

- The key length (which determines the block length) it supports,
- The word size and
- The nominal number of rounds.

The notable ones are TinySec [5], Flexisec [10] and MiniSec[6]. These link layer security protocols have an open-ended design so as to enable the use of any block ciphers with appropriate mode of operation. Also, the range of applications for which the WSNs can be used is very wide. Hence, in order to optimize the security levels- desired vs. resource-consumption trade-off, the link layer security protocol employed must be configurable with respect to (a) the actual cipher and the mode of operation to be employed and, (b) The security attributes desired such as encryption, message authentication or replay protection.

We believe that the efficiency of the block cipher is one of the important factors in leveraging the performance of the link layer protocol. Even though the Skipjack (80-bit cipher key with 64-bit block size) [11] is the default block cipher used by TinySec and MiniSec; We have attempted to carefully investigate the applicability of block cipher Mistyl (128-bit cipher key with 64-bit block size) [12] and OFB against the Cipher Block Chaining (CBC) [13] mode as the desired Block cipher mode of operation.

In this paper, therefore, we present our study and analysis results which **"The Misty cipher in the OFB mode"** over Rijndael, Skipjack cipher in CBC mode, as the baseline to evaluate.

We believe that the actual cipher to use with a specific mode of operation to be employed must be arrived at only looking after and considering the specific security demands of the applications, rather than by following any abstract model. Therefore, for the resource constrained issue in WSN environment **"The Misty cipher in the OFB mode"** implementation and evaluation exercise could be useful in arriving at the choice of the block cipher to be in tune with the available resources and the type of the security desired.

TinySec employs link layer encryption with Skipjack as the default cipher with Cipher Block Chaining (CBC) mode and CBC-MAC (Message Authentication Code) [13]. TinySec exploit the advantage of implementing link layer security in software by providing minimal configurable security attributes. The configurable security allows different modes of operations and support for (a) encryption and authentication (b) only message authentication (provided by default) or (c) disabling the security support altogether. Thus it does not support the configurable link layer security.

MiniSec that is designed for the Telos motes [14]. It uses a different approach in that it offers two operating modes. It offers all the basic desired link layer security properties viz. data encryption, message integrity and replay protection.

In general, the block ciphers used for evaluation in WSN environment are RC5 [15], Rijndael [16], MISTY1 [18]. There have been many benchmarks and evaluation of the block ciphers for the WSNs as surveyed here. But none of them focus specifically on the security at the link layer framework. The evaluation is based on security properties, storage and energy efficiency of the ciphers. The results prescribe Skipjack (low security at low memory), MISTY1 (higher security at low memory) and Rijndael (highest speed but higher memory) as the most suitable ciphers depending upon the availability of memory and the required level of security. Another attempt at evaluating the security mechanisms in WSNs measurements like (a) the impact of packet overhead on energy consumption (b) the impact of different ciphers on the CPU and memory usage (c) the impact of security layer (including cipher) on the message/network throughput, on the network latency and on the energy consumption. We believe that the size of the cipher key is an indicative measure of the strength of the computational security of the cipher. At the minimum, the cipher key size must be enough, so as to prevent the brute force attack against the cipher. With the rapid advancement in technology, the conventional key size of 80-bits is longer sufficient. As per the claims of RSA Security Labs, 80-bit keys would become crack able by 2010 [17]. Hence, it is essential to move towards ciphers with 128-bit cipher key sizes. Our selection is the encryption algorithm MISTY1 is a 64-bit block cipher with 128-bit key MISTY1 was designed on the basis of the theory of "provable security" against differential and linear cryptanalysis. It was also carefully designed to withstand various cryptanalytic attacks that were known at the time of designing.MISTY1 working fine with OFB mode of operations so we select OFB mode of operation.

A MISTY blocks cipher with OFB mode of operation should be used to implement link layer security scheme. To encrypt message longer than block size, a mode of operation is used, also these operation modes do not only affect the security, but also the energy-efficiency of the encryption scheme. The fault tolerance against cipher text error and synchronization error (where bit must be added or lost) must also be taken into account.

Secret-key cryptosystems MISTY1, which is a block ciphers with a 128-bit key, a 64-bit block and a variable number of rounds. MISTY is a generic name for MISTY1. Designed on the basis of provable security theory, against differential and linear cryptanalysis, and moreover they realize high speed encryption on hardware platforms as well as on software environments. With software implementation eight rounds can encrypt a data stream in CBC mode at a speed of 20Mbps and 40Mbps on Pentium/100MHz and PA-7200/120MHz, respectively. For its hardware performance, a prototype LSI by a process of 0.5CMOS gate-array and confirmed a speed of 450 Mbps. MISTY1 can encrypt a data stream in OFB mode with faster speed than CBC mode.

MISTY is to offer secret-key cryptosystems that are applicable to various practical systems as widely as possible, for example software stored in IC cards and hardware used in fast ATM networks. To realize this, following three fundamental policies are followed:

- MISTY should have a numerical basis for its security
- MISTY should be reasonably fast in software on any processor
- MISTY should be sufficiently fast in hardware implementation.

### 3.1 Security analysis

We have chosen to evaluate MISTY1 becauseMISTY1 is one of the CRYPTREC recommended 64-bit ciphers. MISTY1 is designed for high-speed implementations on hardware as well as software platforms by using only logical operations and table lookups. We have also found MISTY1 to be particularly suitable for 16-bit platforms. It is a royalty-free open standard documented in RFC2994. Security Babbage and Frisch demonstrate the possibility of a 7th order differential cryptanalytic attack on 5-round MISTY1, none of the S-boxes with optimal linear and differential properties has an optimal behavior with respect to higher order differential cryptanalysis, however as improvement, the number of rounds of the FI function could be increased. MISTY1 has four instead of three S-boxes in its FI function. Mr.K¨uhn found an impossible differential attack on 4-round MISTY1 using 238 chosen plaintexts and 262 encryptions. In the same paper, a collision search attack has also been described – the attack requires 228 chosen plaintexts and 276 encryptions. In a later paper, Mr.K¨uhn shows that the FL function introduces a subtle weakness in 4-round MISTY1. This weakness allows him to launch a slicing attack with as few as 222.5 chosen plaintexts on 4-round MISTY1. The best known attack on 5-round MISTY1 so far is Knudsen and Wagner's integral cryptanalysis [19], at a cost of 234 chosen plaintexts and 248 time complexity.

Skipjack is 80 bit key and 64 bit block cipher with 32 rounds. The cipher key size must be enough, so as to prevent brute force attack against the cipher. Skipjack is low security at low memory. The major limitation of TinySec is its limited platform support; the official release of TinySec as included in version 1 of TinyOS only works on the MICA2 mote. The current release of MiniSec also suffers from some implementation flaws which artificially limit the length of the 802.15.4 packets by overloading the length field. Furthermore MiniSec overloads a bit in the packet header that CC2420 datasheet specifies as reserved and should always be set to zero. This can lead to unpredictable operation if MiniSec requires this field to be set to 1. In terms of encryption/decryption Rijndael offers the highest speed at the expense of relatively large code size and data memory requirement.

To conclude, the security margin provided by MISTY1 is relatively small compared with modern 128-bit ciphers like Rijndael, but with nominal rounds MISTY1 is still reasonably secure.

### 3.2 Efficiency analysis and experiments

In this section we show one Example of hardware implementation of MISTY1 with eight rounds.

Hardware

We have also designed a prototype LSI of MISTY1 with eight rounds, which has the following specifications:

| Encryption speed | 450 mbps |
|---|---|
| clock | 14 MHz |
| I/O(plain,cipher ,key) | 32 bit parallel_3 |
| Supported Modes | ECB,CBC,OFB-64,CFB |
| Design process | 0.5μ CMOS gate-array |
| Number of gates | 65K gates |
| Package | 208-pin at package |

This LSI has no repetition structure that is; it contains the full hardware of eight FO functions and ten FL functions. It takes two cycles to encrypt a 64-bit plaintext; it also has three independent 64-bit registers that store a plaintext, an intermediate text after the fourth round, and a cipher text, respectively. This structure makes the following pipeline data processing possible:

| cycles | Plain text1 | Plain text2 | Plain text3 | Plain text4 |
|---|---|---|---|---|
| Cycles 1 & 2 | Input | Encryption | Input | |
| Cycles 3 & 4 | | Encryption | Input | |
| Cycles 5 & 6 | Output | Output | Encryption | Input |
| Cycles 7 & 8 | Input | Encryption | Input | |

### 3.3 Size and speed analysis

**Memory**

We refer to two types of memory: (1) code memory, in the form of Flash memory and (2) data memory, in the form of RAM. The memory organization of MSP430F149 is such that data memory ranges from address 0200h to 09FFh, where as code memory ranges from 01100h to 0FFFFh. There are three types of segments for this platform: CODE, CONST and DATA. A typical policy is such that CODE and CONST segments are put in the code memory, while DATA segments (CSTACK, DATA16 AN, DATA16 I, DATA16 N, DATA16 Z and HEAP) in the data memory. For the memory usage of CODE and CONST segments, we can read it off the memory map listing generated by the compiler. For the memory usage of DATA segments except CSTACK, we can also read it off the memory map listing. For CSTACK however, we have to rely on manual inspection of the code, and our estimation is conservative. The storage for plaintext, cipher text, cipher key as well as the expanded key is not included in the code memory or data memory in Table 1. The module named 'skey' in Table 1 and other tables hence forth Refers to the key expansion algorithm The code memory for an operation mode module, say a CBC module, takes into account the code memory for the bare bone encryption code, the CBC code as well as the S-boxes (or lookup tables in general).

**CPU Cycles**

The computational complexity of an algorithm translates directly to its energy Consumption. Assuming the energy per CPU cycle is fixed, and then by measuring the number of CPU cycles executed per byte, we get the amount of energy consumed by byte. For example, the processor used, MSP430F149 draws a nominal current of 420μA at 3V and at a clock frequency of 1MHz in active mode – this means that the energy per instruction cycle (for the processor alone) is theoretically 1.26 nJ. The results can be found in Table 2. Observe that among all modes; only CBC is asymmetrical in the sense that encryption and decryption consume a slightly different number of CPU cycles. From Table 1 and 3, we can see that OFB is not only the most space-saving but also the fastest mode.

**Table 1:** Memory Requirements in bytes.

| Cipher | Module | Size optimized | | Speed optimized | |
|---|---|---|---|---|---|
| | | Code Memory | Data Memory | Code Memory | Data memory |
| RC5-32 | CBC | 1653 | 52 | 6045 | 52 |
| | OFB | 997 | 36 | 5097 | 36 |
| RC6-32 | CBC | 2121 | 86 | 2329 | 86 |
| | OFB | 1740 | 54 | 1834 | 54 |
| Rijndael | CBC | 13448 | 92 | 13998 | 92 |
| | OFB | 12816 | 60 | 13110 | 60 |
| MISTY | CBC | 6973 | 58 | 7969 | 58 |
| | OFB | 6311 | 42 | 7015 | 42 |

**Table 2:** CPU cycles for key expansion (Per key) and operation Mode (per byte).

| Cipher | Module | Size optimized | | Speed optimized | |
|---|---|---|---|---|---|
| | | Encryption | Decryption | Encryption | Decryption |
| RC5-32 | CBC | 1247 | 1270 | 712 | 699 |
| | OFB | 1225 | 1225 | 668 | 668 |
| RC6-32 | CBC | 1222 | 1247 | 1132 | 1132 |
| | OFB | 1205 | 1205 | 1120 | 1120 |
| Rijndael | CBC | 321 | 331 | 218 | 223 |
| | OFB | 299 | 299 | 204 | 204 |
| MISTY | CBC | 478 | 485 | 525 | 531 |
| | OFB | 445 | 455 | 503 | 503 |

With MISTY cipher CBC consumes 478 cpu cycles for encryption and MISTY with OFB consumes 445 cpu cycles for encryption. If we use MISTY with CBC then code memory required 6973 and data memory required 58, but with OFB code and data memory requirement is less than CBC mode with MISTY cipher. So in term of speed optimization also MISTY with OFB mode of operation is a better option.

In terms of key setup, MISTY1 is an attractive option with the least data memory requirement and CPU cycles, and modest code memory requirement. In terms of encryption/decryption, Rijndael offers the highest speed at the expense of relatively large code size and data memory requirement.

MISTY1 is a solid performer in all categories, never falling below the 3rd place, from these two tables we can see that MISTY takes very low code memory compare to Rijndael and also in case of data memory MISTY takes first place for size optimized. If size optimized then sensor node size can be small in shape and also we can get high security with low memory compare to Rijndael and skipjack.

**Table 3:** CPU cycles for Key expansion (Per Key) and Operation Mode (Per byte).

By Key Setup

| Rank | Size optimized | | | Speed optimized | | |
|---|---|---|---|---|---|---|
| | Code Memory | Data Memory | Speed | Code Memory | Data Memory | Speed |
| 1 | RC5-32 | MISTY | MISTY | RC6-32 | MISTY | MISTY |
| 2 | RC6-32 | Rijndael | Rijndael | RC5-32 | Rijndael | Rijndael |
| 3 | MISTY | RC6-32 | RC5-32 | MISTY | RC6-32 | RC5-32 |
| 4 | Rijndael | RC5-32 | RC6-32 | Rijndael | RC5-32 | RC6-32 |

By Key Encryption

| Rank | Size optimized | | | Speed optimized | | |
|---|---|---|---|---|---|---|
| | Code Memory | Data Memory | Speed | Code Memory | Data Memory | Speed |
| 1 | RC5-32 | RC5-32 | Rijndael | RC6-32 | RC5-32 | Rijndael |
| 2 | RC6-32 | MISTY | MISTY | RC5-32 | MISTY | MISTY |
| 3 | MISTY | RC6-32 | RC5-32 | MISTY | RC6-32 | RC5-32 |
| 4 | Rijndael | Rijndael | RC5-32 | Rijndael | Rijndael | RC6-32 |

## 4. Discussion

Cryptographic algorithms are an essential part of the security architecture of WSNs, using the most efficient and sufficiently secure algorithm is thus an effective means of conserving resources. By 'efficient' in this paper we mean requiring little storage and consuming little energy. Data transmission consumes more energy than computation. MISTY1 depending on the combination of available memory with required security is good for storage and energy efficiency. OFB is not only the most space-saving but also the fastest mode. Some sensor network only needs data integrity, availability, confidentiality and semantic type security. These types of sensor networks only needs to reduce power, storage and cost, as well as with a good security. If we can implement block cipher in hardware then speed will increase also, so after consider all these factors we propose an efficient size optimized, low cost, scalable and secured schemes that can make primarily privacy, integrity and availability of packets in wireless sensor network.

By scalable we mean that, MISTY is one of the earliest block ciphers that were aimed at being suitable for both software and hardware system, particularly for low cost applications. It was designed so that in real world encryption performance could be achieved even in heavy resource constrains.

By Low cost we meant, MISTY can implemented with a very small RAM size (e.g. within 100 bytes) with software in cheap processor and with a small number of gate counts and low power consumption (e.g. within 10 gates) in hardware.

A large number of sensor nodes are deployed and the MAC scheme must establish the communication link between the sensor nodes. We present the challenges in the design of the energy efficient medium access control (MAC) protocols for the wireless sensor network. These Wireless Sensor Networks have severe resource constrains and energy conservation is very essential. The sensor node's radio in the WSNs consumes a significant amount of energy. Substantial research has been done on the design of low power electronic devices in order to reduce energy consumption of these sensor nodes. Because of hardware limitations further energy efficiency can be achieved through the design of energy efficient communication protocols.

We propose an efficient link layer security scheme in this paper. It is efficient in cipher computation. By using the MISTY-OFB mode encryption mechanism and the IV techniques proposed scheme takes very low memory size and also code size, MISTY cipher works very fast with OFB mode of operation so it can significantly decrease sensor node size which is of great important to wireless sensor networks like environmental application. MISTY-OFB takes very small space but provides security like confidentiality, data integrity, data freshness and primary privacy of a network which are important concern for environmental WSN network. Sensor network is application specific, size and cost optimization with required security service is our work.

For example, a typical environmental or habitat monitoring applications like (a) tracking the movement of an animal in a sanctuary or (b) monitoring the amount of rainfall in the catchments areas of a river across a dam, to enable forecasting the probability of rainfall downstream, the confidentiality of data packets transmitted to the base station is not essential but the data integrity, entity authentication, message freshness and replay protection are very vital for the applications.

## 5. Conclusion and future work

About the operation mode to use, Apart from having the highest memory and energy efficiency, OFB has desirable fault-tolerance characteristics, i.e. a cipher text error affects only the corresponding bit(s) of the plaintext. Therefore in an error-prone environment such as the wireless network, OFB is particularly useful. However in the event of loss of synchronization, OFB has to rely on an external mechanism to regain synchronization.

For truly constrained environments, we recommend MISTY1. Although MISTY1 is slower than Rijndael, in terms of key setup, MISTY1 uses less memory and CPU cycles than Rijndael does. Furthermore, MISTY1 uses less RAM for encryption, not to mention that the overall code size of MISTY1 is half of Rijndael's. The only drawback of MISTY1 is that it is (believed to be) less secure than Rijndael and it only caters for one key length, but we foresee that there are applications that do not require a security level higher than MISTY1 can provide.

In conclusion, we have presented taking into account the security properties, storage- and energy-efficiency of a set of candidates; we have arrived at a systematically justifiable selection of block ciphers and operation modes. Instead of taking for granted prevalent suggestions, we now have results that we can use not only for convincing ourselves but also other researchers in the area of WSN.

Experiments show that the proposed scheme is more efficient than TinySec. Future work is to implement this scheme with TOSSIM simulator to compare energy-efficiency, power consumption and cost size with other security work like TinySec, Minisec, Sensec, and Flexisec. This work is more effective for environmental type sensing network.

International Workshop, FSE '97. Volume 1267 of LNCS. Springer-Verlag (1997) 54–68, 1997

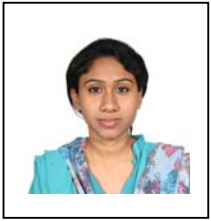

**Nasrin Sultana** earned her B.Sc in Computer Engineering in 2003 and M.Sc. in Computer Science & Engineering in 2010 from American international University – Bangladesh (AIUB). Currently she is working as a Lecturer in Department of Computer Science in the same university. She has taken ccna and ccna security certification. Her research interests include wireless sensor networks, network security, grid computing and mobile & multimedia networks.

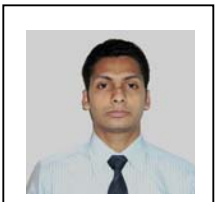

**Tanvir Ahmed** earned his B.Sc. in Computer Science & Engineering in 2008 and M.Sc. in Computer Science in 2009 from the American International University - Bangladesh (AIUB). Currently he is working as a Lecturer in Department of Computer Science in the same university. His research interests include Database Management system, Data Mining, Data security, Embedded System and grid computing.

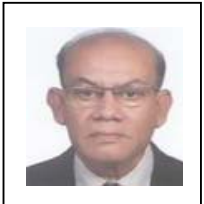

**A B M Siddique Hossain** is currently Professor and Dean, Faculty of Engineering of American International University-Bangladesh (AIUB) since 1$^{st}$ April 2008. He has more than 40 years of teaching, research and Administrative experience as Head of EEE department, Dean of EEE Faculty, Director of Computer center both at home in Bangladesh University of Engineering and Technology (BUET), Dhaka and abraod. He worked as a visiting Professor in a number of Universities at home and abroad. He is a senior member of IEEE, Life member of Institute of Engineers – Bangladesh (IEB) and Bangladesh Computer Society (BCS). He was the Chairman of IEEE Bangladesh Section during the years 1997 to 2000 and received IEEE millennium award. His field of research interest is electronics, computers and wireless communications 3G and Beyond, LTE (4G) etc. He has more than 50 publications in national, international journals and conference proceedings.